%% file: main.tex
\documentclass[10pt,conference]{IEEEtran}
\usepackage[utf8]{inputenc}
\usepackage{enumitem}
\usepackage{graphicx}
\usepackage[flushleft]{threeparttable}
\usepackage{fancybox}
\usepackage{multirow, adjustbox, chngpage}
\usepackage{amssymb}
\usepackage{amsmath, bm, amsfonts}
\usepackage{bigstrut}
\usepackage{color}

\newcommand{\Answer}[1]{\begin{center}%
    \noindent\thicklines\setlength{\fboxsep}{5pt}%
    \cornersize{0.1}\Ovalbox{\begin{minipage}{8.3cm}%
    \textit{#1}\end{minipage}} \end{center}}

\newcommand{\NAME}{CAME}
\newcommand{\RQone}{To what extent historical values of source code metrics can improve detection performances?}
\newcommand{\RQtwo}{How does \NAME{} compare to other static ML algorithms?}
\newcommand{\RQthree}{How does \NAME{} compare to existing detection techniques?}

\begin{document}
\title{Deep Learning Anti-patterns from Code Metrics History}

\author{
    \IEEEauthorblockN{
        Antoine Barbez\IEEEauthorrefmark{1}, Foutse Khomh\IEEEauthorrefmark{1}, Yann-Gaël Guéhéneuc\IEEEauthorrefmark{2}
    }
    \IEEEauthorblockA{
        \IEEEauthorrefmark{1}Polytechnique Montreal, Canada; antoine.barbez@polymtl.ca, foutse.khomh@polymtl.ca\\
        \IEEEauthorrefmark{2}Concordia University, Canada; yann-gael.gueheneuc@concordia.ca
    }
}

\maketitle

\begin{abstract}
Anti-patterns are poor solutions to recurring design problems. Number of empirical studies have highlighted the negative impact of anti-patterns on software maintenance which motivated the development of various detection techniques. Most of these approaches rely on structural metrics of software systems to identify affected components while others exploit historical information by analyzing co-changes occurring between code components. By relying solely on one aspect of software systems (i.e., structural or historical), existing approaches miss some precious information which limits their performances.

In this paper, we propose \NAME{} (Convolutional Analysis of code Metrics Evolution), a deep-learning based approach that relies on both structural and historical information to detect anti-patterns. Our approach exploits historical values of structural code metrics mined from version control systems and uses a Convolutional Neural Network classifier to infer the presence of anti-patterns from this information. We experiment our approach for the widely know God Class anti-pattern and evaluate its performances on three software systems. With the results of our study, we show that: (1) using historical values of source code metrics allows to increase the precision; (2) \NAME{} outperforms existing static machine-learning classifiers; and (3) \NAME{} outperforms existing detection tools.
    
\end{abstract}


\input{01Introduction}
\input{02RelatedWork}
\input{03Approach}

\input{04StudyDesign}
\input{05Study1}
\input{06Study2}

\input{07Threats}
\input{08Conclusion}

\bibliographystyle{plain}
\bibliography{references}

\end{document}

%% file: 01Introduction.tex
\section{Introduction}
During its development and maintenance, a software system may experience what Cunningham~\cite{cunningham1993wycash} called the \textit{technical debt}, i.e., immature code and design choices implemented with the aim of meeting a deadline. The technical debt often takes the form of \textit{anti-patterns}, i.e., poor solutions to recurring design problems, originally proposed by Fowler~\cite{fowler1999refactoring} and Brown et al.~\cite{brown1998antipatterns} along with refactoring operations aimed at removing them. There exists a variety of anti-patterns, which have been shown to negatively impact the maintainability~\cite{yamashita2013exploring} and comprehensibility~\cite{abbes2011empirical} of the code. For example, the God Class anti-pattern, which happens when a class grows rapidly with the addition of new functionalities, violates the principle of single responsibility, which results in a class with low cohesion and high coupling.  

A variety of approaches have been proposed to detect the occurrences of anti-patterns in source code~\cite{Moha10-TSE-DECOR, tsantalis2009identification, marinescu2010incode}. Most of them rely on the formal definitions of anti-patterns and attempt to identify their occurrences in source code using structural metrics (e.g., Lines Of Code) along with empirically defined thresholds. However, anti-patterns can also be detected by an analysis of change history information~\cite{PalombaBPOLP13,Palomba15}. Indeed, the presence of anti-patterns in a system influence how source code entities evolve with one another over time. For example, the Feature Envy anti-pattern, which happens when a method is implemented in the wrong class, can be detected by identifying methods that change more often with methods of another class than those of their own class.

Although structural and historical anti-patterns detection techniques have shown acceptable performances, they are still far from perfect; they often report large numbers of false positives and false negatives. Also, we observe a complementarity in their detection results, meaning that some occurrences can not be detected by approaches based solely on structural or historical information~\cite{PalombaBPOLP13}. Thus, existing approaches miss some precious complementary information which limits their performances. On the one hand, structural detection techniques rely on one single version of software systems. On the other hand, the historical detection technique does not consider the structural properties of the changed entities, nor the nature of the changes they have undergone.

Based on such considerations, we propose \NAME{} (\textbf{C}onvolutional \textbf{A}nalysis of code \textbf{M}etrics \textbf{E}volution), a deep-learning based approach to detect anti-patterns by analyzing how source code metrics evolve over time when changes are applied to the system. Hence, the main idea behind our approach is to exploit the ability of deep-neural networks to identify key features in raw data with the aim of detecting anti-patterns from both structural and historical information. Concretely, structural metrics values related to the code components to be classified are computed at each revision of the system under investigation by mining its version control system (e.g., Git, SVN). This information is then organized into a two dimensional vector and fed through a Convolutional Neural Network (CNN) architecture to perform classification.

To the best of our knowledge, we are the first to exploit structural and historical information simultaneously for detecting anti-patterns.

This paper experiments the proposed approach for the detection of God Class. To train and assess the performances of our model, as well as to compare it with competing approaches, we used a manually-produced oracle containing occurrences of the studied anti-pattern in eight open-source java projects. To verify that our approach leverages historical information about code metrics to increase its performances, we firstly carried out an experiment in which we compare the performances of our model with different lengths of history (i.e., number of revisions). This, study answers the following research question:

\begin{enumerate}[leftmargin=1.1cm]
\item[\textbf{(RQ1)}] \textbf{\RQone{}} \\
Our results indicate that for God Class detection, the performances of our model significantly increase with the length of the input metrics history. With a metrics history of 500 revisions, the F-measure achieved by our model on the three test systems increases by $33\% \pm 6\%$ with respect to the performances achieved using one single revision.
\end{enumerate}

Afterward, we compare the performances of \NAME{} with those achieved by several \textit{static} Machine Learning (ML) classifiers. Here and in the remainder of this paper, the term \textit{static} refers to approaches that do not exploit historical information and therefore rely only on the current revision of the considered system, i.e., the revision in which we want to detect anti-patterns. Thus, we answer the following research question:

\begin{enumerate}[leftmargin=1.1cm]
    \item[\textbf{(RQ2)}] \textbf{\RQtwo{}} \\
    For God Class detection, our results show an overall improvement of 38\% in term of F-measure, with respect to the classifier that achieved the best performances.
\end{enumerate}

Finally, we evaluate our approach as a potential alternative to existing tools for helping practitioners to identify affected components to be refactored. To do so, we compare the performances of \NAME{} with those of three state-of-the-art detection techniques: two static code analysis techniques and one approach that exploits change history information. With the results of this experiment, we aim to answer our last research question:

\begin{enumerate}[leftmargin=1.1cm]
    \item[\textbf{(RQ3)}] \textbf{\RQthree{}} \\
    \NAME{} significantly outperforms existing approaches in detecting the God Class anti-pattern with an F-measure of 0.77. We show that it improves the precision by 196\% and the recall by 51\% with respect to the best competing technique. This suggests that our approach should be considered by practitioners for their software maintenance tasks. 
\end{enumerate}

The remainder of this paper is organized as follows. Section~\ref{section: related work} provides background and discusses the related work. Section~\ref{section: approach} describes our approach \NAME{}. Section~\ref{section: study design} describes our case study design aiming to evaluate our approach for God Class detection. Section~\ref{section: study1 results} reports the results of our first study aiming to assess the impact of the history length on \NAME{}'s performances, thus answering our first research question. Section~\ref{section: study2 results} reports the results obtained while comparing our approach with other techniques and answers the last two research questions. Section~\ref{section: threats} discusses the threats that could affect the validity of our results. Finally, Section~\ref{section: conclusion} concludes with future work.

%% file: 02RelatedWork.tex
\section{Background and Related Work}
\label{section: related work}

\subsection{Definition of God Class}
The God Class anti-pattern refers to the situation in which a class grows rapidly with the addition of new functionalities. A God Class implements a high number of responsibilities, delegating only trivial operations and accessing the data of many other classes. Consequently, this anti-pattern violates the design principle of uniform distribution of the system's intelligence among top-level classes~\cite{riel1996object} and has been shown to decrease program comprehension \cite{abbes2011empirical} and reusability. 

\subsection{Anti-patterns detection}

The negative impact of anti-patterns on software quality highlighted by number of empirical studies~\cite{deligiannis2004controlled, yamashita2012code, yamashita2013exploring, abbes2011empirical} has motivated the development of various automatic detection approaches. Most of these approaches attempt to identify bad motifs in models of source code using manually-defined metric-based heuristics.

First, Marinescu \cite{marinescu2004detection} proposed a mechanism for analyzing source code models with the aim of detecting design defects. This mechanism relies on a set of metrics (along with empirically defined thresholds) logically combined using AND/OR operators. Originally implemented for God Class, this approach called \textit{detection strategy} has later been extended to 11 anti-patterns by Lanza and Marinescu~\cite{lanza2007object} and implemented inside Eclipse plug-ins such as \textit{InCode} \cite{marinescu2010incode}. Similarly, Moha et al.~\cite{Moha10-TSE-DECOR} proposed DECOR (DEtection and CORrection of Design Flaws). DECOR provides detection algorithms for 
four design anti-patterns (God Class, Functional  Decomposition, Spaghetti Code, and Swiss Army Knife) and their 15 underlying code smells. This approach relies on so-called ``Rule Cards'' that encode the formal definitions of anti-patterns and code smells.

Anti-patterns are often defined along with refactoring operations designed to remove them, e.g., a God Class can be removed using the Extract Class Refactoring which consists in splitting the considered class into several more cohesive smaller classes. Consequently, occurrences of an anti-pattern can be detected by identifying opportunities to apply its corresponding refactoring operation. Based on such consideration, Fokaefs et al.~\cite{fokaefs2012identification} proposed an approach to detect God Classes in a system by suggesting a set of Extract Class Refactoring operations. Similarly, Tsantalis and Chatzigeorgiou~\cite{tsantalis2009identification} proposed an approach for automatic suggestions of Move Method Refactoring, i.e., methods that can potentially be moved to another class are considered as potential Feature Envy methods. These heuristics are implemented in the Eclipse plug-in \textit{JDeodorant} \cite{fokaefs2007jdeodorant,fokaefs2011jdeodorant}.

Anti-patterns also impact how code components evolve with one another over time, when changes are applied to the system. Consequently, Palomba et al.~\cite{Palomba15,PalombaBPOLP13} proposed HIST (Historical Information for Smell deTection), an approach to detect anti-patterns by analyzing co-changes occurring between code components. They experimented HIST on eight software systems, for the detection of five anti-patterns: Divergent Change, Shotgun Surgery, Parallel Inheritance, God Class and Feature Envy. They show that HIST is able to identify code smells that cannot be identified through approaches solely based on code analysis and suggest that better performances can be achieved by combining historical and structural information.

\subsection{ML-based Anti-patterns detection}
A number of approaches have 
used ML algorithms to detect anti-patterns. These models also rely on software metrics computed for each component to be classified.
First, Kreimer~\cite{kreimer2005adaptive} relied on decision trees to detect God Class and Long Method. This approach has later been extended to 12 anti-patterns by Amorim et al.~\cite{amorim2015experience}.
Khomh et al.~\cite{khomh2009bayesian, khomh2011bdtex} presented BDTEX (Bayesian Detection Expert), a metric based approach to build Bayesian Belief Networks from the definitions of anti-patterns. This approach has been experimented for the detection of God Class, Functional Decomposition, and Spaghetti Code.
Maiga et al.~\cite{maiga2012smurf,maiga2012support} proposed the use of Support Vector Machines to detect four well known anti-patterns: God Class, Functional Decomposition, Spaghetti code, and Swiss Army Knife.

Fontana et al.~\cite{fontana2016comparing} conducted a large-scale study on the effectiveness of machine learning algorithms for anti-patterns detection. They experimented 16 different machine learning algorithms along with boosting techniques for the detection of four anti-patterns (Data Class, God Class, Feature Envy, and Long Method) on 74 software systems belonging to the \texttt{Qualitas Corpus} dataset \cite{tempero2010qualitas}. They conclude that the application of machine learning to the detection of anti-patterns can provide high accuracy, even with a limited number of training examples.

More recently, Liu et al.~\cite{liu2018deep} proposed a deep learning based approach to detect Feature Envy. Their model is a CNN fed with structural metrics as well as lexical information. They evaluated their approach on seven open-source applications. To train their model, they also propose an automatic approach for generating Feature Envy examples.

\subsection{Anti-patterns Through Systems' History}

Olbrich et al.~\cite{olbrich2009evolution} studied the impact of anti-patterns on the change behavior of code components. Specifically, they analyzed the historical data of two large scale software systems and compared the change frequency and size of components affected by God Class and Shotgun Surgery with those of healthy components. Vaucher et al.~\cite{vaucher2009tracking} studied the ``life cycle'' of God Class occurrences in two open-source systems with the aim of understanding when they are created and how they evolve. They used a Bayesian Belief Network to track the evolution of the``godliness'' of these classes through different versions of the studied systems. Similarly, Chatzigeorgiou and Manakos~\cite{chatzigeorgiou2010investigating} tracked the evolution of three anti-patterns: Long Method, Feature Envy and State Checking in the history of two open-source systems. Finally, Tufano et al.~\cite{tufano2015and} performed the largest experiment on the presence of anti-patterns through the history of software systems. Specifically, they mined the history of 200 software projects to understand under what circumstances anti-patterns appear. First, their results confirm the observation made by Chatzigeorgiou and Manakos~\cite{chatzigeorgiou2010investigating} that most of instances are introduced when the file is added to the system. Second, they show that anti-patterns are also often introduced the last month before deadlines by experienced developers. 

%% file: 03Approach.tex
\section{\NAME{}: Convolutional Analysis of code Metrics Evolution}
\label{section: approach}

This section presents \NAME{}, our deep-learning based approach to detect anti-patterns from source code metrics historical information. We first describe the input of our model before describing its architecture. Finally, we discuss the procedure we followed to train this architecture for anti-patterns detection. 

\subsection{Input}
To detect instances of a given anti-pattern in a system, our approach uses a deep-learning based classifier to perform a boolean prediction on each code component (i.e., class or method) of the system. To perform this prediction, we exploit both structural and historical information. To define the input of our model, we first select a set of $N_{m}$ structural metrics which can be computed for any code component of the system under investigation. Then, we walk through the history of revisions of the system in reverse order by mining its repository. At each revision, we recalculate the values of the selected metrics for each component. We refer to the value of the $j^{th}$ metric computed for a component $c$ at the $i^{th}$ revision of the system, as $m_{i, j}(c) \in \mathbb{R}$. Thus, for a given code component $c$ to be classified, the input of our model is a real valued matrix $\textbf{X}_{c}$ such as:

\begin{equation}
    \label{input matrix}
    \textbf{X}_{c}(i,j) = m_{i,j}(c)
\end{equation}

Different software systems experience a different number of revisions (i.e., commits). Also, a code component may have been introduced in a system during its creation or on the contrary during a recent revision. As a consequence, different code components may be characterized by a different number of revisions. To allow our model to receive a fixed size input, we limit the length of the metrics history (i.e., number of revisions considered for a given component) to a constant $L_{h}$, which is an hyper-parameter to be adjusted. Hence, each input matrix of our model is of shape: $L_{h} \times N_{m}$. If a component has an history shorter than $L_{h}$, meaning that the component did not exist in the early revisions of the system, we pad the rest of its input matrix with zeros. 

\subsection{Architecture}
\label{section: architecture}

\begin{figure}
\centering
\includegraphics[width=\linewidth]{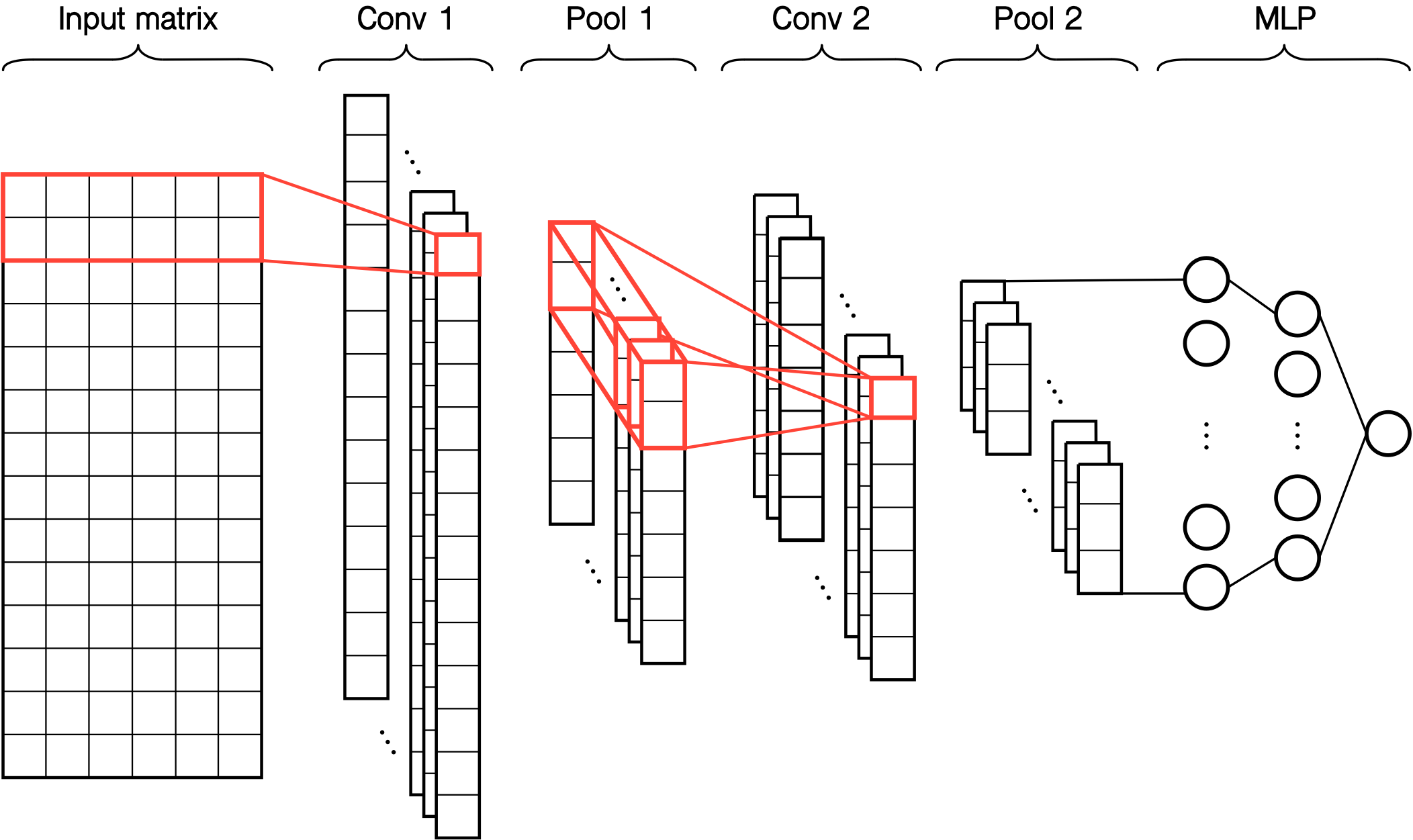}
\caption{Architecture of the CNN model used by \NAME{}. In this example, the network contains two convolution + polling layers with filters of size 2 and two dense hidden layers.}
\label{fig: came architecture}
\end{figure}

Figure~\ref{fig: came architecture} overviews the architecture of the convolutional neural network used by our approach to perform classification.
The model contains several convolution + pooling layers fully-connected to a Multi Layer Perceptron (MLP) model, i.e., several dense layers connected to an output layer. The convolution layers perform 1D convolutions with a \textit{stride} of \textit{1} and a \textit{tanh} activation function. The output of each convolution layers is fed to max-pooling layers that reduce the dimentionality of their input by taking the maximum value across a small spatial region. Then, after being flattened, the output of the last pooling layer is fed to \textit{tanh} dense layers. Finally, the output layer is made of one single \textit{sigmoid} neuron which outputs the predicted probability that a component is affected given its input matrix.

The choice of such architecture has been motivated by the ability of CNNs to extract high level features from high dimensional data, e.g., deep CNNs used in image processing recognize complex shapes in raw pixels. Indeed, our conjecture is that the first convolution layer can detect that some metrics have changed between two consecutive commits. Then, the next layers build a representation of the changes that characterize an anti-pattern. Therefore, our model identify recurrent patterns in the local variations of software metrics, which we believe constitute a useful complementary information for detecting anti-patterns. 

\subsection{Training}
This subsection presents the consideration we adopted to train our model but we assume that the following applies to anyone willing to train neural-networks on the task of anti-patterns detection. First, let $\mathcal{D} = \{(\textbf{X}_{i}, y_{i})\}_{i=1}^{n}$ be our training set. With $\textbf{X}_{i} \in \mathbb{R}^{L_{h} \times N_{m}}$, the input matrix (cf. equation~\ref{input matrix}) of the $i^{th}$ component to classify, $y_{i} \in \{0, 1\}$ the true label for this component and $n$ the number of instances in the training set.

Also, we refer to the set of weights of our model as $\bm{\theta} = \{\textbf{w}_{l}\}_{l=1}^{L}$, with $\textbf{w}_{l}$ being the weight matrix of the $l^{th}$ layer and $L$ the number of layers in the network. Finally, we note $P_{\bm{\theta}}(1|\textbf{X}_{i}) = \text{g}(\textbf{X}_{i}.\bm{\theta})$ the predicted probability outputed by our model for the $i^{th}$ component, with $\text{g}(x) = \frac{1}{1 + e^{-x}}$ the \textit{sigmoid} function. 

\subsubsection{Loss Function}
In a software system, components affected by an anti-pattern are usually a minority ($\approx 1\%$)~\cite{palomba2018diffuseness}. Classifiers optimized using conventional loss functions, e.g., \textit{cross entropy}, on such data tend to favor the majority category, thus maximizing the overall accuracy~\cite{he2008learning}. This characteristic known as the ``imbalanced data problem'' prevents us from using such loss function to guide the optimization of our model. Hence, we must define a loss function that maximizes our evaluation metric: the F-measure expressed in equation~\ref{f-measure}.

Optimizing a model through gradient decent requires computing the gradient of the loss with respect to the model weights. However, computing the number of true positives $TP$ and positives $m_{pos}$ (cf. Table~\ref{Table: confusion matrix}) requires counting elements from the probability outputed by the model, which necessarily involves discontinuous operators:
\begin{equation}
TP(\bm{\theta}, \mathcal{D}) = \sum_{\substack{i=1 \\ y_{i} = +1}}^{n}  \delta(P_{\bm{\theta}}(1|\textbf{X}_{i}) > 0.5)
\end{equation}
\begin{equation}
m_{pos}(\bm{\theta}, \mathcal{D}) = \sum_{i=1}^{n} \delta(P_{\bm{\theta}}(1|\textbf{X}_{i}) > 0.5)
\end{equation}

With:

\[\delta(x) = \bigg\{\begin{tabular}{l}
1 if $x$=True\\
0 if $x$=False
\end{tabular}\]

This characteristic prevents us from using the F-measure directly to define our loss function. Consequently, we use the continuous and differentiable approximation of the F-measure provided by Jansche~\cite{jansche2005maximum} which consists in considering the limit:

\begin{equation}
    \delta(P_{\bm{\theta}}(1|\textbf{x}_{i}) > 0.5) = \lim_{\gamma \to + \infty} \text{g}(\gamma\textbf{X}_{i}.\bm{\theta})
\end{equation}

Hence, we define our loss function as $loss = -\tilde{F}_{m}(\bm{\theta}, \mathcal{D})$ with:

\begin{equation}
    \tilde{F}_{m}(\bm{\theta}, \mathcal{D}) = 2 \times \frac{\sum_{\substack{i=1 \\ y_{i} = +1}}^{n}  \text{g}(\gamma\textbf{X}_{i}.\bm{\theta})}{n_{pos} + \sum_{i=1}^{n} \text{g}(\gamma\textbf{X}_{i}.\bm{\theta})}
\end{equation}
Note that the value of the hyper-parameter $\gamma$ will be adjusted along with other hyper-parameters during the tuning of our model. 

\subsubsection{Regularization}
Overfitting occurs when a statistical model fails to generalize to new examples by learning irrelevant characteristics of its training data. Hence, an overfitted model performs well on the training set but achieves poor performances on unseen test data. To prevent our model from overfitting, we used the widely adopted $L_{2}$ regularization technique, which consists in adding a term to the loss function to encourage the weights to be small \cite{witten2016data}. This term rely on the Euclidean norm of the weight matrices, i.e., $\|\textbf{w}\|_{2}=\sqrt{\textbf{w}^{\top}\textbf{w}}$, also called $L_{2}$-norm. Thus, the $L_{2}$ regularization term added to the loss function can be expressed as:

\begin{equation}
L_{2} = \lambda \sum_{l=1}^{L+1} \|\textbf{w}_{l}\|_{2}
\end{equation}

With $\lambda \in \mathbb{R}$ an hyper-parameter adjusted during cross-validation.

%% file: 04StudyDesign.tex
\section{Study Design for God Class}
\label{section: study design}
In this section, we lay the foundations of our study aiming to evaluate the effectiveness of our approach \NAME{} for detecting the God Class anti-pattern. After describing the software systems investigated in this work, we present the metrics selected for God Class and overview the process followed to extract historical information about these metrics. We finally describe our evaluation approach and replication package. Our study aims at addressing the following three research questions:
\begin{enumerate}[leftmargin=1.1cm]
    \item[\textbf{(RQ1)}] \textbf{\RQone{}} \\
    This research question assesses the impact of the length of the input metrics history (i.e., $L_{h}$) on our models' performances. Hence, before comparing our approach with other detection techniques, we can confirm that metrics history provides relevant information to our model and establish which value of $L_{h}$ leads to optimal performances. 
    
    \item[\textbf{(RQ2)}] \textbf{\RQtwo{}} \\
    This research question aims at comparing the performances of our approach with other static ML algorithms, i.e., which rely on one single revision of the systems.
    
    \item[\textbf{(RQ3)}] \textbf{\RQthree{}} \\
    This research question aims at comparing our approach with existing detection techniques. The results of this comparison will provide insights on the advantages of using \NAME{} instead of other detection tools to help developers in their daily maintenance tasks.
\end{enumerate}

\subsection{Studied Systems}
\label{subsection: studied systems}
To answer our research questions, we base our study on eight open-source Java projects belonging to various ecosystems. Android Opt Telephony and Android Support belong to the Android APIs\footnote{https://android.googlesource.com/}. Apache Ant, Apache Tomcat, Apache Lucene, and
Apache Xerces belong to the Apache Foundation\footnote{https://www.apache.org/}. ArgoUML\footnote{http://argouml.tigris.org/} is a software design tool and Jedit\footnote{http://www.jedit.org/} a text editor. For the sake of simplicity, we chose to analyze only the directories that implement the core features of the systems and to ignore test directories.

The choice of these systems has been motivated by the fact that they have been used for a similar purpose in prior studies. Indeed, to train our model and compare its performances with those of other approaches, we needed an oracle reporting the occurrences of God Classes in a set of software systems. Unfortunately, we found no such large dataset in the literature.
Consequently, we reused the occurrences of God Class used to evaluate the approaches HIST~\cite{PalombaBPOLP13} and DECOR~\cite{Moha10-TSE-DECOR}, made publicly available\footnote{ http://www.rcost.unisannio.it/mdipenta/papers/ase2013/}\footnote{http://www.ptidej.net/tools/designsmells/materials/} by their respective authors. Among the systems available, we kept only those for which the full history was available through Git or SVN. Also, for some systems, we found occurrences that do not belong to it or that do not exist in the current revision. In such cases, we did not incorporate the systems in our oracle. Table~\ref{Table: systems} overviews the main characteristics of the subject systems.

\begin{table}
\caption{Characteristics of the Studied Systems}
\label{Table: systems}
\centering
\begin{tabular} { l l l l l}
\hline
System name & Snapshot & Directory & \#Files & \#GCs \\ \hline
Android Opt Telephony&c241cad&src/java/&190&10 \\
Android Support&38fc0cf&v4/&104&4 \\
Apache Ant&e7734de&src/main/&755&7\\
Apache Lucene&39f6dc1&src/java/&160&3\\
Apache Tomcat&398ca7ee&java/org/&1005&5\\
Apache Xerces&c986230&src/&658&15\\
ArgoUML&6edc166&src\_new/&1246&22 \\
Jedit&e343491& ./&437&5\\ \hline
\end{tabular}
\end{table}

\subsection{Source Code Metrics}
\label{subsection: source code metrics}
For God Class detection, the components to classify correspond to the classes of the system. However, we chose to consider only top-level-classes as potential God Classes, as we found no inner-classes positively labeled in our data. To decide whether or not a given class is affected by the God Class anti-pattern, we retrieve the history of \textbf{seven} structural metrics. Note that to compute these metrics, we do not consider attributes and methods of inner- (or nested-) classes as components of the class under investigation. Indeed, the refactoring operation commonly applied to remove God Classes (Extract Class Refactoring) consists in identifying one or several group of attributes and methods of the class dedicated to one functionality and then extract them as a separate class. Thus, we assume that inner-classes may be the result of such refactoring. In the following, we define each selected metric and provide a brief rational for their use.

\begin{itemize}
    \item \textbf{ATFD} (Access To Foreign Data): Number of distinct attributes of unrelated classes (i.e., not inner- or super-classes) accessed (directly or via accessor methods) in the body of a class. A God Class accesses a lot of data from other classes as suggested by Lanza and Marinescu~\cite{lanza2007object}.
    \item \textbf{LCOM5} (Lack of COhesion in Methods): Measures cohesion among methods of a class based on the attributes accessed by each method~\cite{henderson1995object}. A God Class handles many unrelated functionalities, thus methods related to different functionalities access different sets of attributes. This metrics is also part of the detection process of DECOR~\cite{Moha10-TSE-DECOR}. 
    \item \textbf{LOC} (Lines Of Code): Sum of the number of lines of code of all methods of a class. A God Class implements a high number of functionalities, thus it is mainly characterized by its size.
    \item \textbf{NAD} (Number of Attributes Declared): Number of attributes declared in the body of a class. This metrics is part of the detection process of DECOR~\cite{Moha10-TSE-DECOR}.
    \item \textbf{NADC} (Number of Associated Data Classes): Number of dependencies with data-classes (i.e., data holders without complex functionality other that providing access to their data). This metrics is part of the detection process of DECOR~\cite{Moha10-TSE-DECOR}.
    \item \textbf{NMD} (Number of Methods Declared): Number of non-constructor and non-accessor methods declared in the body of a class. This metrics is part of the detection process of DECOR~\cite{Moha10-TSE-DECOR}.
    \item \textbf{WMC} (Weighted Method Count): Sum of the McCabe's cyclomatic complexity~\cite{mccabe1976complexity} of all methods of a class. A God Class has a high functional complexity, as suggested by Lanza and Marinescu~\cite{lanza2007object}. 
\end{itemize}

\subsection{Data Extraction}
As previously evoked, we construct the input matrices of our model by mining software repositories and navigating through the different revisions of a system using its version control API (e.g., Git). To automate this process and allow replication, extension and reuse of our work, we designed a component called \texttt{RepositoryMiner}\footnote{https://github.com/antoineBarbez/RepositoryMiner/} which automatically extracts the source code metrics history of any java software system. The \texttt{RepositoryMiner} currently implements 12 class- and method-related structural metrics and allows to easily define new ones. We briefly describe each step of the process followed to extract our data below.

\noindent\textbf{Input:}
The \texttt{RepositoryMiner} takes as input three arguments: (1) the URL of the system's repository; (2) the SHA, i.e., the identification number, of the system's snapshot (i.e., commit) we want to analyze and; (3) the sub-directories of interest, i.e., those in which we want to detect affected components. 

\noindent\textbf{Initialization}
After downloading the system repository, we checkout to the current snapshot and parse the Abstract Syntax Tree (AST) of all the \textit{.java} files contained in the sub-directories of interest. This step creates a \texttt{SystemObject} that holds a representation of the system's classes which will be used later to compute our metrics.

\noindent\textbf{Data Extraction}
We walk through all the commits of the system in reverse order starting from the current snapshot. At each commit, we perform the following steps:
\begin{enumerate}
\item Retrieve the names of all the files that have been changed between the current and the previous revision.
\item If the changed files belong to the \texttt{SystemObject}, checkout and update the corresponding classes.
\item Check if any component (i.e., file, class, or method) have been renamed between the current and the next revision.
\item Compute the code metrics values for each class of the \texttt{SystemObject} by taking into account renamed components.
\end{enumerate}

\noindent\textbf{Output}
The \texttt{RepositoryMiner} outputs a collection of \textit{.csv} metric files. Each file contains the code metrics values computed for each class of the system at a given revision. Note that before creating a new metric file, we check if at least one value has been modified with respect to the previous one.

\subsection{Evaluation}
\label{subsection: evaluation}
To evaluate the performances of \NAME{} as well as those of the competing classifiers, we selected three systems, i.e., Android Support, Apache Tomcat, and Jedit, among the eight software systems presented in Table~\ref{Table: systems}. These systems have been selected for the sake of generalizability. Indeed, they belong to different domains: telephony framework, service container, and text editor and have different sizes (i.e., number of classes). The remaining five systems are used for training and tuning the hyper-parameters of the different ML-based approaches investigated in this work. 

We compute the overall performances of each approach by running it on all instances (i.e., the java classes) of the three test systems. Indeed, each classifier is able to perform a boolean prediction on each single instance. Then, we evaluate a classifier using the so-produced confusion matrix presented in Table~\ref{Table: confusion matrix}. 

\begin{table}[htb]
\caption{Confusion Matrix for Binary Classification}
\label{Table: confusion matrix}
\begin{center}
\renewcommand{\arraystretch}{2}
\begin{tabular} {c  c | c  c | c }
& & \multicolumn{2}{ c |}{\textbf{\textit{predicted label}}} & \multirow{2}{*}{\rotatebox[origin=c]{90}{\textbf{\textit{total}}}}\\ 
& & $1$ & $0$ & \\ \hline
\multirow{2}{*}{\rotatebox[origin=c]{90}{\textbf{\textit{true label}}}}& $1$ & $TP$ & $FN$ & $n_{pos}$ \\
& $0$ & $FP$ & $TN$ & $n_{neg}$\\ \hline
\multicolumn{2}{ c |}{\textbf{\textit{total}}}& $m_{pos}$& $m_{neg}$ & $n$
\end{tabular}
\end{center}
\end{table}

With $TP$ the number of true positives; $FN$ the number of false negatives (i.e., misses); $FP$ the number of false positives and $TN$ the number of true negatives. We use this matrix to compute our evaluation metrics:

\begin{center}
\begin{minipage}{.5\linewidth}
\begin{equation}
  precision = \frac{TP}{TP + FP}
\end{equation}
\end{minipage}%
\begin{minipage}{.5\linewidth}
\begin{equation}
  recall = \frac{TP}{TP + FN}
\end{equation}
\end{minipage}
\end{center}

In order to evaluate each approach with a single aggregated metric, we also compute the F-measure (i.e., the harmonic mean of precision and recall):

\begin{equation}
\label{f-measure}
F_{m} = 2\times\frac{precision \times recall}{precision + recall} = 2\times\frac{TP}{n_{pos} + m_{pos}}
\end{equation}

\subsection{Replication Package}
To facilitate further evaluation and reuse of our work, all the data used in the context of this study, as well as our implementation of \NAME{} for God Class is publicly available in our online appendix\footnote{https://github.com/antoineBarbez/CAME/}.

%% file: 05Study1.tex
\section{Assessing the Impact of the Metrics History Length on \NAME{}'s Performances}
\label{section: study1 results}

In this section, we report the results of our experiments conducted with the aim of assessing the impact of the metrics history length ($L_{h}$) on the performances of our approach. Hence, this section answers the first research question.

\subsection{Approach}
\label{subsection: study1 approach}
To answer \textbf{RQ1}, we monitor the performances achieved by \NAME{}, in terms of precision, recall and F-measure, with different lengths of metrics history: $L_{h} \in \{1, 10, 50, 100, 250, 500, 1000\}$. For each value of $L_{h}$ experimented, we build and train 10 distinct CNNs in order to retrieve the mean and standard deviation of the three performance metrics achieved for each length. To avoid any bias in our conclusions, we must consider that the length of the metrics history may affect the optimal values of the other hyper-parameters of our model. Consequently, before training our model for a new value of $L_{h}$, we perform a new tuning of its hyper-parameters. As explained in Section~\ref{subsection: evaluation} the performances metrics are computed by considering instances of the three test systems together: Apache Tomcat, Jedit and Android Support. The remaining five systems are kept for training and hyper-parameters tuning. 

\subsection{Hyper-parameters Tuning}
\label{subsection: study1 tuning}
We select the optimal set of hyper-parameters of our model for each history length investigated using a random search over 100 generations of nine hyper-parameters~\cite{bergstra2012random}. Table~\ref{Table: came tuning} reports for each hyper-parameter, the range of values experimented. We monitor the performances achieved by our model with different sets of hyper-parameters by carrying out a 5-fold cross-validation. Thus, we first split the training set into 5 equal size partitions, i.e., folds. Then, at each iteration, we generate a new random set of hyper-parameters and compute our model's prediction on each fold by leaving it out while keeping the others for training (100 epochs). Finally, we concatenate the obtained predictions and compute the overall F-measure. The optimal values retained after tuning for each history length can be found in our online appendix\footnote{https://github.com/antoineBarbez/CAME/tree/master/experiments/tuning/}. 

\begin{table}
\caption{\NAME{} Hyper-parameters Tuning} 
\label{Table: came tuning}
\centering
\begin{threeparttable}
\begin{tabular} { l l }
\hline
Hyper-parameter & Range \\ \hline
Learning Rate ($\eta$) & $10^{-[0.0; 2.5]}$ \\
L2-norm ($\lambda$) &  $10^{-[0.0; 2.5]}$ \\
Gamma ($\gamma$) & $[1; 10]$ \\
\# Conv Layers & $[0, 1]$ if $L_{h}\leq10$ else $[1; 2]$ if $L_{h}\leq100$ else $2$ \\
\# Filters & $[10; 60]$ \\
Filter size & $[2; 4]$ \\
Pool size & $\{2, 5, 10\}$ if $L_{h}\leq100$ else $\{5, 10, 15, 20\}$ \\
\# Dense Layers & $[1; 3]$ \\
Dense Layer size & $[4; 100]$ then $[4; s]$ \\ \hline
\end{tabular}
\begin{tablenotes}
\small
\item \textit{With $s$ the size of the previous dense layer.}
\end{tablenotes}
\end{threeparttable}
\end{table}

For each history length investigated, we train our model using the previously found hyper-parameter values. We perform a mini-batch based stochastic gradient descent optimization during $300$ epochs, with $5$ mini-batches and an exponential learning rate decay of $.5$ every $100$ epochs. It is important to remember that we train $10$ randomly initialized CNNs per history length investigated in order to compute the mean and standard deviation of their performance metrics.

\subsection{Results}
\begin{figure}
\centering
\includegraphics[width=9.2cm]{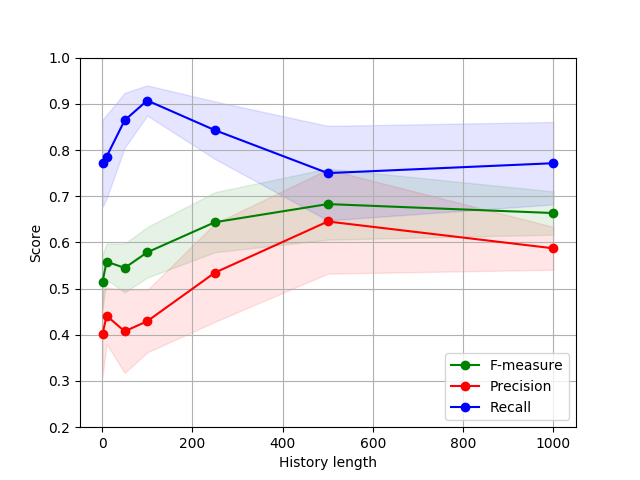}
\caption{Comparison of \NAME{}'s performances for different sizes of metrics history. The lines show mean values while areas show standard deviations over 10 trainings.}
\label{fig: came_compare_hist_length}
\end{figure}

Figure~\ref{fig: came_compare_hist_length} presents the results of our comparison for God Class, i.e., mean values and standard deviations of \NAME{}'s performance metrics over the three test systems, obtained using different sizes of metrics history (1, 10, 50, 100, 250, 500, 1000).

\vspace{.2cm}
\noindent\textit{RQ1: \RQone{}}
\vspace{.15cm}

For God Class detection, our results show that in term of F-measure, the overall performances achieved by \NAME{} on the three test systems significantly increase with the size of the metrics history. Specifically, after training our model ten times per history length, we observe that the overall F-measure improves from $0.51 \pm 0.06$ for $L_{h}=1$ to $0.68 \pm 0.08$ for $L_{h}=500$ (improvement of $33\% \pm 6\%$). Regarding the other two performance metrics, we see that although \NAME{} achieves a better recall on average for $L_{h} > 1$, there does not seem to be any real correlation between recall and history length. As shown in Figure~\ref{fig: came_compare_hist_length} the recall strongly increases in the range $[1;100]$ but then drops to recover its initial value at $L_{h}=1000$. Finally, we can see that the precision clearly improves with the history length similarly to the F-measure. By comparing the precision for $L_{h}=1$ with respect to the history length that led to the best results ($L_{h}=500$), we observe an overall improvement of $61\% \pm 11\%$.

Also, we can see that the performances of \NAME{} (especially precision) decrease for $L_{h}>500$. This observation is surprising considering that a longer metrics history includes the shorter ones and thus should lead to a greater or equal F-measure. We believe this could be due to a sub-optimal choice of the hyper-parameters. Indeed, the randomness of the process and the finite number of combinations experimented make it hard to conclude that the hyper-parameters selected are optimal. Furthermore, while tuning our model, we limited the number of convolution layers to 2. Hence, it is possible that more convolution layers may be needed to process a history of size $L_{h}=1000$ and achieve optimal performances.

\Answer{Our results confirm that our model properly leverages historical information about source code metrics by decreasing the number of false positives. Specifically, for $L_{h}>1$ we observe a recall greater or equal to that obtained with a single revision of the input metrics and more importantly, we see that the precision clearly increases with the length of the input metrics history.}

%% file: 06Study2.tex
\section{Comparing \NAME{} with Other Approaches}
\label{section: study2 results}

This section reports the results of our experiments aiming to compare \NAME{} with other approaches. To avoid redundancies, we report the results for our last two research questions together.

\subsection{Approach}
Similarly to the previous study, we evaluate the performances of the different approaches on our three test systems (i.e., Apache Tomcat, Jedit, and Android Support) while keeping the other five for training and hyper-parameters calibration. To compute the performances of \NAME{} on these systems, we reuse the ten CNNs trained during our previous study (cf. Sections \ref{subsection: study1 approach}) with $L_{h} = 500$. As shown in Figure~\ref{fig: came_compare_hist_length}, after being trained, each CNN achieves different performances due to the randomness of its initialization. Hence, the final performances of \NAME{} are computed from the outputs of the ten CNNs using an ensemble method, also known as \textit{boosting technique}. Such method have been shown to lead to better performances than those of each independent classifier~\cite{dietterich2000ensemble}.    
In the context of this study, we use the widely-adopted Bayesian averaging heuristic to compute the ensemble prediction. Thus, the final predicted probability that a class $c$ is a God Class can be expressed as:

\begin{equation}
P_{ensemble}(1|\textbf{X}_{c}) = \frac{\sum_{i=1}^{10} P_{i}(1 | \textbf{X}_{c})}{10}
\end{equation} 

\noindent with $P_{i}(1 | \textbf{X}_{c})$, the predicted conditional probability given by the $i^{th}$ CNN and $\textbf{X}_{c}$ the input matrix of the class $c$.

To answer \textbf{RQ2}, we compare \NAME{} with three ML classifiers: Decision Tree, Multi Layers Perceptron (MLP), and Support Vector Machine (SVM). The input of these classifiers consists in the same seven metrics used by \NAME{} (cf. Section~\ref{subsection: source code metrics}) computed on the current revision of each studied system. Similarly to \NAME{}, we first calibrate the hyper-parameters of each classifier and compute the final performances using the Bayesian averaging ensemble method on ten pre-trained models.

To answer \textbf{RQ3}, we compare \NAME{} with three state-of-the-art detection approaches. Two static code analysis techniques: DECOR~\cite{Moha10-TSE-DECOR} and JDeodorant~\cite{fokaefs2011jdeodorant} and one approach that exploits change history information to detect anti-patterns: HIST~\cite{PalombaBPOLP13}. We chose to compare \NAME{} with these approaches to increase the scope of our study. Indeed, they rely on radically different strategies to detect anti-patterns and are thus likely to be complementary. DECOR rely on the use of \textit{Rule Cards} that encode the formal definitions of anti-patterns using structural and lexical information. JDeodorant detects affected components by identifying refactoring opportunities. Finally, HIST exploits change history information derived from version control systems. To compute the performances of each approach,  we used the implementations made publicly-available by their respective authors whenever possible and replicated the approaches for which no implementation was available. Thus, we ran DECOR using the Ptidej API\footnote{https://github.com/ptidejteam/v5.2/} and JDeodorant using its Eclipse plug-in\footnote{https://marketplace.eclipse.org/content/jdeodorant/}. For HIST, we implemented the detection rules as described in its original paper \cite{PalombaBPOLP13}. Also, we implemented our own component\footnote{https://github.com/antoineBarbez/HistoryExtractor} to extract code changes at a class level granularity because the original component was not available due to its license.

\subsection{Hyper-parameters Tuning}

\begin{table}
\caption{Hyper-parameters Tuning of the Competing ML Classifiers} 
\label{Table: classifiers tuning}
\centering
\begin{threeparttable}
\begin{tabular} { c l l }
\hline
Model & Hyper-parameter & Range \\ \hline
\multirow{4}{*}{Decision Tree} & Max Features & $\{sqrt, log2, None\}$\\
& Max Depth & $10\times[1; 10]$\\
& Min Sample Leaf & $\{1, 2, 4, 6\}$\\
& Min Sample Split & $\{2, 5, 10, 15\}$\\ \hline
\multirow{5}{*}{MLP} & Learning Rate ($\eta$) & $10^{-[0.0; 2.5]}$ \\
& L2-norm ($\lambda$) &  $10^{-[0.0; 2.5]}$ \\
& Gamma ($\gamma$) & $[1; 10]$ \\
& \# Dense Layers & $[1; 3]$ \\
& Dense Layer size & $[4; 100]$ then $[4; s]$ \\ \hline
\multirow{3}{*}{SVM} & Penalty & $\{0.001, 0.01, 0.1, 1, 10, 100\}$\\
& Gamma & $\{0.001, 0.01, 0.1, 1, 10, 100\}$\\
& Kernel & $\{linear, rbf, sigmoid\}$\\ \hline
\end{tabular}
\begin{tablenotes}
\small
\item \textit{With $s$ the size of the previous dense layer.}
\end{tablenotes}
\end{threeparttable}
\end{table}

We calibrate the hyper-parameters of each ML algorithm using the same procedure adopted for \NAME{} in the previous study (see Section~\ref{subsection: study1 tuning}). Hence, we use a random search of 100 iterations over a variety of hyper-parameters related to each classifier. Also, to monitor the performances induced by each set of hyper-parameters, we use a 5-fold cross-validation with instances of the five training systems. Table~\ref{Table: classifiers tuning} reports for each classifier, the set of hyper-parameters tuned, as well as the ranges of values experimented.
To train the MLP model, we used rigorously the same procedure followed for training \NAME{}.

\begin{table*}
\caption{Performances for God Class detection}
\label{Table: performances god class}
\begin{adjustbox}{width=\textwidth}
\renewcommand{\arraystretch}{1.2}
\begin{tabular}{|c|c|c|c|c|c|c|c|c|c|c|c|c|}
\hline
\multirow{2}{*}{Approaches}& 
\multicolumn{3}{c|}{
	Apache Tomcat
} 
&\multicolumn{3}{c|}{
	JEdit
}
&\multicolumn{3}{c|}{
	Android Platform Support
}
&\multicolumn{3}{c|}{
	\textbf{Overall}
}\bigstrut [t] \\ 
\cline{2-13}
&\textit{Precision}&\textit{Recall}&\textit{F-measure}
&\textit{Precision}&\textit{Recall}&\textit{F-measure}
&\textit{Precision}&\textit{Recall}&\textit{F-measure}
&\textit{Precision}&\textit{Recall}&\textit{F-measure} \bigstrut [t]\\
\hline
DECOR &68\%&40\%&50\%&17\%&60\%&26\%&--&0\%&--&24\%&36\%&29\% \bigstrut \\ \hline
HIST &--&0\%&--&25\%&40\%&31\%&18\%&100\%&31\%&20\%&43\%&27\% \bigstrut \\ \hline
JDeodorant &2\%&60\%&5\%&5\%&60\%&9\%&50\%&50\%&50\%&4\%&57\%&8\% \bigstrut \\ \hline \hline
Decision Tree &33\%&20\%&25\%&100\%&40\%&57\%&100\%&25\%&40\%&68\%&29\%&40\% \bigstrut \\ \hline
MLP &25\%&100\%&40\%&68\%&80\%&73\%&100\%&75\%&86\%&41\%&\textbf{86\%}&56\% \bigstrut \\ \hline
SVM &50\%&20\%&29\%&100\%&20\%&33\%&--&0\%&--&68\%&14\%&24\% \bigstrut \\ \hline \hline
\textbf{\NAME{}} &50\%&100\%&68\%&100\%&80\%&89\%&100\%&75\%&86\%&\textbf{71\%}&\textbf{86\%}&\textbf{77\%} \bigstrut \\ \hline
\end{tabular}
\end{adjustbox}
\end{table*}

\subsection{Results}

Table~\ref{Table: performances god class} reports the performances for God Class detection, in terms of precision, recall and F-measure, achieved by \NAME{} along with those of the competing techniques. The performances are reported for each subject system and on overall, i.e., considering the three test systems as a single one. When an approach did not detect any occurrence of God Class in a system, it was possible to compute neither the precision, nor the F-measure. In these cases a “--” is indicated in the corresponding cell.

\vspace{.2cm}
\noindent\textit{RQ2: \RQtwo}
\vspace{.15cm}

For God Class detection, our results show that on overall, our approach significantly outperforms the three classifiers. In term of F-measure the performances improve from 56\% to 77\% (improvement of 38\%) with respect to the classifier with the best performances (the MLP). Unsurprisingly, \NAME{} achieves the same recall than the MLP. Indeed, it is interesting to remark that a MLP model is equivalent to the CNN used by \NAME{} with no convolution layers and considering $L_{h} = 1$. Hence, we make the same observation than in the previous study regarding recall. Considering the other two classifiers we clearly see that they can not compete with \NAME{} in term of recall with 29\% for the Decision Tree algorithm and 14\% for the SVM against 86\% for our approach. Finally, \NAME{} ensured on overall a better precision than any other classifier (+ 18\% on average). Regarding the performances achieved on each system, we see that \NAME{} achieves the highest precision (100\%) on two of the three test systems but only 50\% on Apache Tomcat. This may be due to the fact that this system is the largest of our training set but contains only five occurrences of God Class. Hence, any model is more likely to have a low precision on it. Finally, our approach seem to have a more stable recall with values ranging between 75\% and 100\%.

\Answer{\NAME{} significantly outperforms other static ML classifiers. Indeed, none of the competing algorithms performs better than our approach on any system and considering any performance metric.}

\vspace{.2cm}
\noindent\textit{RQ3: \RQthree}
\vspace{.15cm}

Our results show that \NAME{} clearly outperforms existing detection methods in detecting the God Class anti-pattern. Indeed, \NAME{} shows on overall, a precision of 71\% and a recall of 86\% (F-measure of 77\%) over the test systems. With respect to the tool that performs the best for each performance metric, we see that \NAME{} improves the precision by 196\%, the recall by 51\% and the F-measure by 166\%. Also, as we can see, each tool achieves poor performances on at least one system, which is not the case for our approach. This confirms that \NAME{} performs well independently of the systems characteristics.

However, some factors can explain the poor performances reported for some of the approaches experimented. First, as the original implementation of HIST is not publicly available, we had to replicate this approach from the directives given by the authors in the paper~\cite{PalombaBPOLP13}. We are aware that some differences in our respective implementations may have impacted the performances reported. Second, it is important to note that JDeodorant is not, strictly speaking, an anti-pattern detection tool. Instead, JDeodorant suggests opportunities to apply refactoring operations in the system. Hence, it is not surprising that it achieves a high recall at the expense of its precision, because it may exists a high number of classes in the subject systems that could benefit from an Extract Class Refactoring operation without necessarily being God Classes. 

\Answer{\NAME{} significantly outperforms all the existing techniques investigated in this work for God Class detection. This suggests that our approach should be considered by practitioners for identifying affected code components to be refactored.}

%% file: 07Threats.tex
\section{Threats to Validity}
\label{section: threats}
In this section, we discuss the threats that could affect the validity of our results.

\paragraph{Construct Validity}
Threats to construct validity concern the relation between theory and observation. This could be due to how our oracle was build (cf. Section~\ref{subsection: studied systems}). To combat this limitation, we examined the occurrences reported in respectively HIST and DECOR's replication packages before incorporating them in our oracle. Also, both papers have been awarded by the community which convince us of the reliability of their data. Also, the metrics we selected for God Class detection have been chosen on the basis of their use in literature. However, other metrics that we have overlooked may have led to better performances. Yet, our approach does not focus on any specific metric but on the potential improvements induced by the use of metrics history in general. Another threat is related to the implementations used to evaluate existing approaches. To replicate HIST, we followed rigorously the guidelines provided by the authors. However, some differences may remain between our respective implementations. 

\paragraph{External Validity} Threats to external validity concern the generalizability of our findings. In the context of our study, this could refer to the number of software systems on which we experimented our approach. Manually-build datasets are time consuming to produce for anti-patterns which explains their rareness in literature. Hence,
our sample size may limit the generalizability of our results and we agree that further evaluation of our approach on a larger set of systems would be desirable. To reduce this threat, the software systems used for evaluation have been selected for their different sizes and domains. This threat could also refer to the fact that we experimented our approach for the detection of one single anti-pattern: God Class. Yet, we assume that our approach could be applied to any anti-pattern and we plan to extend it in a future work.

\paragraph{Internal Validity}
Threats to internal validity concern all the factors that could have impacted our results. This could refer to the fact that the ML classifiers investigated in this work, i.e., \NAME{}, Decision Tree, MLP, SVM, are not deterministic approaches. As a consequence, the predictions made by these classifiers may vary depending on their initialization which makes them difficult to compare. To reduce this threat, we base our evaluation on ten independent models per classifier. Hence we consider their individual performances as a Gaussian distribution in our first study (cf. Section~\ref{section: study1 results}) and use an ensemble method in our second study (cf. Section~\ref{section: study2 results}). Also, we calibrated the hyper-parameters of all the approaches investigated in this work whenever necessary. Finally, Another threat is related to the choice of the ML model used by our approach to process the input metrics history. As explained in Section~\ref{section: architecture}, we chose a CNN model because we believe that applying successive convolutions to the input allows our model to learn a representation of the changes that characterize an anti-pattern. However, this decision was only based on intuition and we cannot exclude that other deep-learning models (e.g., RNNs) would lead to better performances.    

%% file: 08Conclusion.tex
\section{Conclusion and Future Work}
\label{section: conclusion}
The impact of anti-patterns on software quality highlighted by number of empirical studies has motivated the development of various detection techniques. Although the proposed approaches have helped developers in identifying affected code components to be refactored, we identified a major limitation common to these works. Different detection techniques rely on different sources of information and thus, identify different sets of occurrences of anti-patterns. Hence, by ignoring either structural or historical aspects of software systems, existing approaches miss some precious information which limits their performances.
Consequently, we proposed \NAME{} a deep-learning based approach that relies on both structural and historical information to detect anti-patterns. Our approach exploits historical values of structural code metrics and uses a CNN classifier to infer the presence of anti-patterns from this information. We implemented our approach for the detection of God Class and evaluated it on three software systems. With the results of our study, we showed that:
\begin{itemize}
    \item The performances of \NAME{} increase with the length of the metrics history fed through our model. This shows that our model properly leverages historical information to improve its performances.
    \item \NAME{} significantly outperforms other ML-based classifiers that do not rely on historical data.
    \item \NAME{} significantly outperforms existing detection tools.
\end{itemize}
Our short term research agenda includes extending our approach to other anti-patterns as well as further evaluation of \NAME{} on a greater number of systems. We also plan to investigate the use of deep-learning visualization techniques~\cite{zeiler2014visualizing} on the architecture of \NAME{}. We believe that such approach could help us identifying the root causes and characteristics of anti-patterns.